\documentclass[aps,prd,onecolumn,groupedaddress,showpacs,showkeys, nofootinbib,amssymb]{revtex4}
\usepackage[T1]{fontenc}
\usepackage[latin1]{inputenc}
\usepackage{graphicx}
\usepackage[english]{babel}
\usepackage{bm}
\usepackage{amsmath}
\usepackage{amssymb}
\usepackage{amsfonts}
\usepackage{epsfig}
\usepackage{colordvi}
\usepackage{color}

\begin{document}

\title{Noether Symmetry Approach for  teleparallel-curvature cosmology}
\author{Salvatore Capozziello$^{1,2,3}$
Mariafelicia  De Laurentis$^{4}$\footnote{e-mail address: mfdelaurentis@tspu.edu.ru}, Ratbay Myrzakulov$^{5}$
 }
\affiliation{$^{1}$Dipartimento di Fisica, Università
di Napoli {}``Federico II'', Compl. Univ. di
Monte S. Angelo, Edificio G, Via Cinthia, I-80126, Napoli, Italy}.
\affiliation{$^{2}$INFN Sezione  di Napoli, Compl. Univ. di
Monte S. Angelo, Edificio G, Via Cinthia, I-80126, Napoli, Italy.}
\affiliation{$^3$Gran Sasso Science Institute (INFN), Via F. Crispi 7, I-67100, L' Aquila, Italy.}
\affiliation{$^{4}$Tomsk State Pedagogical University, 634061 Tomsk, Russian Federation.}
\affiliation{$^5$Eurasian International Center for Theoretical Physics and Department of General Theoretical Physics, Eurasian National University, Astana 010008, Kazakhstan.}

\date{\today}

\begin{abstract}
We consider curvature-teleparallel $F(R,T)$ gravity, where the gravitational Lagrangian density is given by
an arbitrary function of the Ricci scalar $R$ and the torsion scalar $T$. Using the Noether Symmetry Approach, we  show that the functional form of the $F(R, T)$ function,  can be determined by the presence of symmetries . Furthermore, we obtain exact  solutions through to the presence of conserved quantities and the reduction of cosmological dynamical system. Example of particular cosmological models are considered.  

 \end{abstract}
\maketitle

\section{Introduction}
\label{uno}
In recent years the increased interest for the Extended Theories of Gravity has led to figure out the cosmic acceleration phenomenon (dark energy) under the standard of further gravitational degrees of freedom coming from  generalized gravities \cite{PhysRepnostro,OdintsovPR,5,6,Mauro,faraoni,9,10,libri,libroSV,libroSF}.  This means  that phenomena like dark energy and dark matter could be addressed assuming a different behavior of the gravitational field with respect to the standard General Relativity (GR) at infrared scales \cite{annalen}. In particular,  modified gravitational  theories like $f(R)$-gravity  can be considered as extensions of General Relativity,   alternatives to dark matter and  dark energy.  In these classes of theories, generic functions of the Ricci scalar $R$  are considered, for example, to address the accelerated expansion observed by supernovae observations \cite{supernovae1,supernovae2}.

Many models have been introduced starting from the primitive  $f(R)$ extension \cite{quintessence}, such as for example $F({\cal G})$, where ${\cal G}$  is Gauss-Bonnet topological invariant or combinations of these last two, as the $F(R, {\cal G})$ \cite{F(GR)-gravity,17,18,21,32,antonio,felixquint,double}. Furthermore, extension of teleparallel gravity, $f(T)$, where $T$ is the torsion scalar have been considered \cite{ft}. The general issue is that many geometric  invariant can be considered and the problem to find a new "material component" to address the accelerated expansion problem could be be completely  circumvented assuming extensions of GR. However, the problem is {\it how many} and {\it what kind} of geometric invariants can be used, Besides, what kind of physical information one can derive from them. 

Recently much interest has also been given to the, $F(R,T)$ modified theories of gravity, where the gravitational Lagrangian constituted  by an arbitrary function of the Ricci scalar $R$ and the  torsion scalar  $T$ \cite{ratbay,ratbay1,zerb}.  The problem could seem redundant since information contained in $f(R)$ gravity could be the same contained in $f(T)$ gravity depending on the definition of connection (e.g. Levi-Civita or Weitzenb\"{o}ck). Actually, differences emerge when the theories are reduced under the same standard. In the Friedman-Robertson-Walker (FRW) metric, differences emerge pointing out that degrees of freedom of $f(R)$  and $f(T)$ are not exactly the same \cite{greci}.  This fact emerges when one searches for symmetries of the theories that are, in general, different. 

In this paper, we want  to obtain Noether symmetries for $F(R,T)$ Lagrangians and, consequently, to fix  specific forms of the Lagrangian. The method of  Noether Symmetry Approach has been  extensively used for alternative theories  giving some relevant results both for cosmology and self gravitating-systems \cite{greci,cimento,hamilton,pla,pla2,marek,india,stabile1}.
 The present paper is organized as follow. In Sec. \ref{due} we introduce briefly the theoretical motivations and the main ingredients for   $F(R,T)$ gravity. In particular, we point out how, $R$ and $T$ degrees of freedom can be discussed under the same standard comparing holonomic and anholonomic coordinate systems.
 In Sec. \ref{tre} we derive the FRW cosmological equations for $F(R,T)$ gravity starting from the point-like Lagrangian that can be derived considering the function $R$ and $T$ as suitable Lagrange multipliers. In such a case, $R$ and $T$ can be considered as independent fields. 
The Noether Symmetry Approach  is discussed in Sec. \ref{quattro}. The existence of the symmetries allows to fix the form of the $F(R,T)$ function and to find out exact cosmological solutions. 
Conclusions  are drawn in Sec. \ref{cinque}.

\section{$F(R,T)$ gravity}
\label{due}
A general torsionless action $f(R)$ gravity is given by
\begin{equation}
{\cal A}=\frac{1}{2\kappa}\int d^4x \sqrt{-g}f(R)+{\cal L}_m 
\end{equation}
with $g$ being the determinant of the metric tensor, ${\cal L}_m$ is the matter part of the action, $\kappa=8\pi G_N$ and $f(R)$ is a non-linear function of curvature scalar $R$.
As we well know,  this Lagrangian can be  obtained directly by replacing the Ricci scalar $R$ in the Hilbert-Einstein Lagrangian with a function $f(R)$ of the Ricci scalar.
On the other hand,  $f(T)$-gravity is the modified form of the curvature-free {\it vierbein} gravitation theory which is also known as the teleparallel gravity.  Following the same method adopted for  $f(R)$-gravity, $f(T)$-gravity can be directly  achieved by replacing the torsion scalar $T$ with a general function of torsion $f(T)$ in the teleparallel Lagrangian  \cite{16a,barrow,Myrzakulov:2010a,Kazuharu:2010a,Kazuharu:2010b,Dent:2010, Li:2010}. The theory is described by the following action
\begin{equation}
{\cal A}=\frac{1}{2\kappa}\int d^4x h f(T)+{\cal L}_m\,, 
\end{equation}
where $h$ is the determinant of the vierbein.
It is important to note that the field equations for $f(T)$-gravity are second order in the covariant derivatives  and therefore  simpler than $f(R)$-gravity that are of fourth order.  
The vierbein $h_i^{\mu}$ has the following properties\footnote{In this study, we represent the space-time indices by the Greek alphabet  $(\alpha, \beta,\mu,\nu...)$
and the tangent space indices by the Latin alphabet ($a,b,i,j...$). These indices run over the values $0,1,2,3$.}
\begin{eqnarray}
h_i^\mu h_\nu^j\ =\ \delta^\mu_\nu, \ \ \ h_i^\mu h^i_\mu\ =\ \delta^\mu_\nu,
\end{eqnarray}
and it is considered like a dynamical object. Here $h_\mu^j$ is the inverse matrix of vierbein. The vierbein relates with the metric as
\begin{eqnarray}\label{eq:metric}
g_{\mu\nu} = \eta_{ab}h^a_\mu h^b_\nu
\end{eqnarray}
where $\eta_{ab}={\rm diag}(1,-1,-1,-1)$ is the Minkowski metric
for the tangent space. 
The action for the theory  where the Lagrangian is a combination of Ricci  and Torsion scalars is
\begin{equation}\label{action1}
 {\cal A}=\frac{1}{2\kappa}\int d^4 x\sqrt{-g}\left[F(R,T)+{\cal L}_m\right]\,.
 \end{equation}
The curvature scalar $R$ is defined by $R=g^{\mu\nu}R_{\mu\nu}$ where $g^{\mu\nu}$ is the inverse of the metric tensor and $R_{\mu\nu}$ is the Ricci tensor. 
Clearly, we can define  $|h|\equiv$ det$\left(h^i_\mu\right)=\sqrt{-g}$ in order to connect the two formalisms.
The Einstein-Hilbert theory is built on the Levi-Civita connection of the metric 
\begin{eqnarray}\label{eq:levi-civita}
\Gamma^\alpha_{\mu\nu} &\equiv& \frac{1}{2}g^{\alpha\beta}\left(\partial _\mu g_{\beta\nu} +\partial_\mu g_{\beta\nu} - \partial_\beta g_{\mu\nu}\right)\,.
\end{eqnarray}
The above connection has non-zero curvature and it is yet torsionless. Using the torsionless Levi-Civita connection the Ricci tensor assume the following form
\begin{eqnarray}
R_{\alpha\beta}=\partial_\eta\Gamma^\eta_{\beta \alpha}-\partial_\beta\Gamma^\eta_{\eta \alpha}+\Gamma^\eta_{\eta\lambda}\Gamma^\lambda_{\beta\alpha}-\Gamma^\eta_{\beta\lambda}\Gamma^\lambda_{\eta\alpha}\,.
\end{eqnarray}
While in the theory of vierbein we use  Weitzenb\"{o}ck connection
(tilded to
distinguish from Levi-Civita connection $\Gamma^\alpha_{\mu\nu}$)
\begin{eqnarray}\label{eq:weizenbock}
\tilde{\Gamma}^\alpha_{\mu\nu} =h^\alpha_\kappa \partial_\nu h^\kappa_\mu\,,
\end{eqnarray}
that has a zero curvature but nonzero torsion. The torsion tensor is
\begin{eqnarray}\label{eq:torsion}
T^\alpha_{\mu\nu} &\equiv& \tilde{\Gamma}^\alpha_{\nu\mu}  - \tilde{\Gamma}^\alpha_{\mu\nu}\,,
\end{eqnarray}
and, the torsion scalar $T$ in the action is given by
\begin{eqnarray}\label{eq:torsionsca}
T=S_\rho^{\mu\nu}T^\rho_{\mu\nu}\,.
\end{eqnarray}
where $S_\rho^{\mu\nu}$ is 
\begin{eqnarray}\label{eq:S}
S^{\rho\mu\nu} &\equiv& K^{\mu\nu\rho} - g^{\mu\nu}T^{\sigma\mu}_{\ \ \ \sigma} + g^{\rho\mu}T^{\sigma\nu}_{\ \ \ \sigma}.
\end{eqnarray}
and 
\begin{eqnarray}\label{eq:S}
K^{\mu\nu}_\rho=\frac{1}{2}\left[T^{\nu\mu}_\rho+T_\rho^{\mu\nu}-T^{\mu\nu}_\rho\right]\,.
\end{eqnarray}
Clearly, defining the relation between holonomic and anholonomic reference frames is possible to reduce all these quantities under the same standard.
Hence, the variation of the action allows to find out the field equations. It is 

\begin{eqnarray}
\delta{\cal A}=\frac{1}{2\kappa}\int
d^4x\left[F(R,T)\delta h+h\delta F(R,T)\right]+\delta{\cal L}_m=0\,,\end{eqnarray}
where $\delta F(R,T)$  can be expanded as
\begin{eqnarray}
h\delta F(R,T)=h \frac{\partial F(R,T)}{\partial R}\delta R+h\frac{\partial F(R,T)}{\partial T}\delta T
\end{eqnarray}
The problem is how we can find a relation between $\delta R$ and $\delta T$ because we must focus on the following integral  
\begin{eqnarray}
{\cal I}=\int d^4x h \delta F(R,T)=\int d^4x \left[h\frac{\partial F(R,T)}{\partial R}\delta R+h\frac{\partial F(R,T)}{\partial T}\delta T
\right]
\end{eqnarray}

It is easy to see that
\begin{eqnarray}\label{dricci}
\delta R=\delta \left(g^{\mu\nu}R_{\mu\nu}\right)=R_{\mu\nu}\delta g^{\mu\nu}+\left( \nabla^\mu\nabla^\nu-g^{\mu\nu}\nabla^\lambda\nabla_\lambda\right)\delta g_{\mu\nu}\,,
\end{eqnarray}
\begin{eqnarray}\label{dtorsion}
h\delta T=h\delta\left(S^{\mu\nu}_\rho T^\rho_{\mu\nu}\right)=\left[2\partial_\nu \left(h h^\rho_\kappa S^{\mu\nu}_\rho \right)-2hh^\gamma_\kappa S^{\rho\beta\mu}T_{\rho\beta\mu}\right]\delta h^\kappa_\mu-2\partial_\nu\left(hh^\rho_\kappa S^{\mu\nu}_\rho\delta h^\kappa_\mu\right)\,,
\end{eqnarray}
Using these results we find, after integration by parts,
\begin{eqnarray}
{\cal I}&=&\int d^4xh\left[\left(\nabla^mu\nabla^\nu-g^{\mu\nu}\nabla^\lambda\nabla_\lambda\right)\frac{\partial F(R,T)}{\partial R}-\frac{\partial F(R,T)}{\partial R}R^{\mu\nu}\right]\delta g_{\mu\nu}+\nonumber\\&&+\int d^4x \left[ 2\frac{\partial F(R,T)}{\partial T}\partial_\nu \left(h h_\kappa^\rho S^{\mu\nu}_\rho\right)-2h \frac{\partial F(R,T)}{\partial T}h^\gamma_\kappa S^{\rho\beta\mu} T_{\rho\beta\mu}+2h\left(\partial_\nu\frac{\partial F(R,T)}{\partial T}\right) h^\rho_\kappa S^{\mu\nu}_\rho\right]\delta h^\kappa_\mu\,.
\end{eqnarray}
At this point, we can use the relation between metric tensor and vierbeins. After,  we can define the variation of the action with respect to the vierbeins, and the following field equations come out:

\begin{eqnarray}\label{field}
&&\frac{1}{2}h^\mu_\kappa F(R,T)+h_{\kappa\nu}\left[\left(\nabla^\mu \nabla^\nu-g^{\mu\nu}\nabla^\lambda\nabla_\lambda\right)\frac{\partial F(R,T)}{\partial R}-\frac{\partial F(R,T)}{\partial R}R^{\mu\nu}\right]+\frac{1}{h}\frac{\partial F(R,T)}{\partial T}\partial_\nu \left(h h^\sigma_\kappa S_\sigma^{\mu\nu}\right)\nonumber\\&&-\frac{\partial F(R,T)}{\partial T}h^\gamma_\kappa S^{\rho\beta\mu}
T_{\rho\beta\gamma}+h^\sigma_\kappa S^{\mu\nu}_\sigma\left(\frac{\partial^2 F(R,T)}{\partial T^2}\partial_\nu T+\frac{\partial^2 F(R,T)}{\partial T\partial R}\partial_\nu R\right)=0\,.
\end{eqnarray}
It is easy to see that from  $F(R,T)$ both $f(T)$ and  $f(R)$ can be immediately   recovered. The Hilbert-Einstein action is immediately recovered for  $F(R,T)=R$.  Now we have all the ingredients to derive the cosmological equations.
\section{$F(R,T)$  cosmology}
\label{tre}
The  cosmological equations can be derived both from the field Eqs.(\ref{field})  or deduced by a  point-like canonical Lagrangian ${\cal L}(a,{\dot a}, R,{\dot R},T,{\dot T})$ related to the action (\ref{action1}).  Here  ${\mathbb Q}\equiv\{a,R,T\}$ is the configuration space from which it is possible to derive 
${\mathbb{TQ}}\equiv \{a,\dot{a}, R, \dot{R}, T,{\dot T}\}$,  the corresponding tangent
space  on which ${\cal L}$ is defined as an application.  The variables $a(t)$,
$R(t)$ and $T(t)$ are, respectively,   the scale factor, the  Ricci scalar and the torsion scalar defined in the FRW metric. 
The Euler-Lagrange equations are
\begin{eqnarray}
\frac{d}{dt}\frac{\partial {\cal L}}{\partial {\dot a}}=\frac{\partial {\cal L}}{\partial  a}\,, \qquad
\frac{d}{dt}\frac{\partial {\cal L}}{\partial {\dot R}}=\frac{\partial {\cal L}}{\partial  R}\,,\qquad
\frac{d}{dt}\frac{\partial {\cal L}}{\partial {\dot T}}=\frac{\partial {\cal L}}{\partial  T}\,,
\label{moto3}
\end{eqnarray}
with the energy condition
\begin{eqnarray}
E_{\cal L}= \frac{\partial {\cal L}}{\partial {\dot a}}{\dot a}+\frac{\partial {\cal L}}{\partial {\dot R}} {\dot R}+\frac{\partial {\cal L}}{\partial {\dot T}} {\dot T}-{\cal L}=0\,.
\label{energy}
\end{eqnarray}
Here the dot indicates the derivatives with respect to the cosmic time $t$.
One can use the method of  Lagrange
multipliers to set  $R$  and $T$ as  constraints for  dynamics \cite{makarenko}. In fact selecting
 suitable Lagrange multipliers  and integrating by parts to eliminate higher order derivatives, the
Lagrangian ${\cal L}$ becomes canonical. In physical units,  the action  is 

\begin{eqnarray}
 {\cal A}=2\pi^2\int dt\,a^3 \left\{  F(R,T)-\lambda_1 \left[R+6\left( \frac{{\ddot a}}{a}+\frac{{\dot a}^2}{a^2}\right)\right]-\lambda_2\left[T+6\left( \frac{{\dot a}^2}{a^2}\right)\right]     \right\}\,.
\end{eqnarray}
Here the definitions of the Ricci scalar and the torsion scalar in FRW metric have been adopted, that is  

 \begin{equation} \label{RH}
R=-6\left[ \frac{{\ddot a}}{a}+\left(\frac{{\dot a}}{a}\right)^2\right]=-6(\dot{H}+2H^2)\,,
\end{equation}

 \begin{equation}\label{TH}
 T=-6\left(\frac{{\dot a}}{a}\right)^2 =-6H^2.
 \end{equation}
where a  spatially flat FRW  spacetime has been adopted. It is worth stressing that the two Lagrange multipliers are comparable but the order of derivative is higher for $R$.
By varying the action with respect to $R$ and $T$, one obtains

\begin{eqnarray}
\lambda_1= \frac{\partial F(R,T)}{\partial  R}\,,\qquad \lambda_2= \frac{\partial F(R,T)}{\partial T}\,,
\end{eqnarray}
then the above action becomes
\begin{eqnarray}
 {\cal A}=2\pi^2\int dt \left\{ a^3  F(R,T)- a^3  \frac{\partial F(R,T)}{\partial  R} \left[R+6\left( \frac{{\ddot a}}{a}+\frac{{\dot a}^2}{a^2}\right)\right]-a^3\frac{\partial F(R,T)}{\partial T}\left[T+6 \left(\frac{{\dot a}^2}{a^2}\right)\right] \right\}\,.
\end{eqnarray}
After an integration by parts, the point-like Lagrangian assumes the following form

\begin{eqnarray}\label{PointLagra}
 {\cal L}&=&a^3\left[ F(R,T)-R \frac{\partial F(R,T)}{\partial  R}-T \frac{\partial F(R,T)}{\partial  T}\right]+6\,a{\dot a}^2 \left[\frac{\partial F(R,T)}{\partial  R}-\frac{\partial F(R,T)}{\partial  T}\right]+\nonumber\\&&6\,a^2\,{\dot a}\,\left[{\dot R}\frac{\partial^2 F(R,T)}{\partial  R^2}+ {\dot T}\frac{\partial^2 F(R,T)}{\partial R \partial T}\right]\,,
\end{eqnarray}

which is a canonical function of 3 coupled fields $a$, $R$ and $T$  depending on time $t$. The first term in square brackets has the role of an effective potential.
It is worth  stressing again  that the Lagrange multipliers have been  chosen by considering the definition of the Ricci curvature scalar $R$ and the torsion scalar $T$. This fact allows us to consider the constrained dynamics as canonical. 

It is interesting to consider some important  subcases of the Lagrangian (\ref{PointLagra}). For  $F(R,T)=R$,  the  GR Lagrangian   is recovered. In this case,  we have
\begin{eqnarray}
 {\cal L}=6a{\dot a}^2 +a^3R\,,
\end{eqnarray}
that, after developing $R$, easily reduces to ${\cal L}=-3a{\dot a}^2$, the standard point-like Lagrangian of FRW cosmology. 
In the case $F(R,T)=f(R)$, we have \cite{PhysRepnostro}
\begin{eqnarray}
 {\cal L}= 6 a {\dot a}^2 f'(R)+ 6 a^2  {\dot a}{\dot R}f''(R)+a^3\left[f(R)-Rf'(R)\right]\,,\label{LfR}
\end{eqnarray}
while teleparallel cosmology \cite{greci} is recovered for  $F(R,T)=f(T)$, and then
\begin{eqnarray}
 {\cal L}= a^3[f(T)-Tf'(T)]-6a{\dot a}^2f'(T)
 \label{LfT}\,.
\end{eqnarray}
Clearly,  these cases deserve a specific investigation.
\subsection{The cosmological  equations}
\label{tre2}

 Let us now derive the Euler-Lagrange equations from Eqs. (\ref{moto3})- (\ref{energy}).  They are 

  \begin{eqnarray} 
 &&\left[ \frac{\partial F(R,T)}{\partial R}-\frac{\partial F(R,T)}{\partial T}\right]\left(12{\dot a}^2-6a^2+12a{\ddot a}\right)-3a^2 \left[F(R,T)-T\frac{\partial F(R,T)}{\partial T}-R\frac{\partial F(R,T)}{\partial R}\right]\nonumber\\&&-12 a {\dot a}\left[{\dot T}\frac{\partial^2 F(R,T)}{\partial T^2}-{\dot R}\frac{\partial^2 F(R,T)}{\partial R^2}\right]-12 a {\dot a}\left[{\dot R}\frac{\partial^2 F(R,T)}{\partial R\partial T}-{\dot T}\frac{\partial^2 F(R,T)}{\partial R\partial T}\right]\nonumber\\&&+6a^2\left[{\ddot T}\frac{\partial^2 F(R,T)}{\partial R\partial T}+{\ddot R}\frac{\partial^2 F(R,T)}{\partial R^2}+{\dot T}^2\frac{\partial^3 F(R,T)}{\partial R\partial T^2}+2{\dot R}{\dot T}\frac{\partial^3 F(R,T)}{\partial R^2\partial T}+{\dot R}^2\frac{\partial^3 F(R,T)}{\partial R^3}\right] =0\,,
\label{moto11}
\end{eqnarray}

\begin{eqnarray} 
&&a^3 \left[R \frac{\partial^2 F(R,T)}{\partial R^2}+T\frac{\partial^2 F(R,T)}{\partial R\partial T}\right]+6a{\dot a }^2\left[\frac{\partial^2 F(R,T)}{\partial R^2}+\frac{\partial^2 F(R,T)}{\partial R\partial T}\right]+6a^2{\ddot a}\frac{\partial^2 F(R,T)}{\partial R^2}=0\,,
\label{moto22}
\end{eqnarray}

  \begin{eqnarray} 
&&a^3 \left[T \frac{\partial^2 F(R,T)}{\partial T^2}+R\frac{\partial^2 F(R,T)}{\partial R\partial T}\right]+6a{\dot a }^2\left[\frac{\partial^2 F(R,T)}{\partial T^2}+\frac{\partial^2 F(R,T)}{\partial R\partial T}\right]+6a^2{\ddot a}\frac{\partial^2 F(R,T)}{\partial R\partial T}=0\,,
\label{moto33}
\end{eqnarray}
The  energy condition  (\ref{energy}),  corresponding to the $00$-Einstein equation, gives

\begin{eqnarray} 
E_{\cal L}&=&6a{\dot a}^2\left[\frac{\partial F(R,T)}{\partial R}-\frac{\partial F(R,T)}{\partial T}+\right]+a^3\left[F(R,T)-T\frac{\partial F(R,T)}{\partial T}-R\frac{\partial F(R,T)}{\partial R}\right]\nonumber\\\nonumber\\&&-6a^2{\dot a}\left[{\dot T}\frac{\partial ^2F(R,T)}{\partial R\partial T}+{\dot R}\frac{\partial^2 F(R,T)}{\partial R^2}\right]=0\,,
\label{energy1}
\end{eqnarray}
Alternatively, this system can be derived from the field Eqs.(\ref{field}).

\section{The Noether Symmetries Approach}
\label{quattro}

The existence of Noether symmetries allows to select constants of motion so that the
dynamics results simplified. Often such a dynamics is exactly solvable by a straightforward
change of variables where acyclic ones are determined \cite{safe}.
A Noether symmetry for the Lagrangian (\ref{PointLagra}) exists if the condition 

\begin{eqnarray} 
L_X {\cal L}\,=\,0 \qquad \rightarrow \qquad X{\cal L}\,=\,0\,,
\label{LX}
\end{eqnarray}
holds. Here $L_X$ is the Lie derivative with respect to the Noether vector $X$. Eq.(\ref{LX}) is
nothing else but the contraction of the Noether vector $X$, defined on the tangent space ${\mathbb{TQ}}\equiv\{a,\dot{a}, R, \dot{R}, T,{\dot T}\}$ of the Lagrangian ${\cal L}={\cal L}(a,{\dot a}, R,{\dot R},T,{\dot T})$, with the Cartan one-form, generically defined as

\begin{eqnarray} 
\theta_{\cal L} \equiv \frac{\partial {\cal L}}{\partial {\dot q}_i}dq^i\,.
\end{eqnarray}
Condition (\ref{LX}) gives 

\begin{eqnarray} 
i_X \theta_{\cal L} = \Sigma_0\,,
\end{eqnarray}
where $i_X$ is the inner derivative and $\Sigma_0$ is the conserved quantity \cite{porco,35,36,fGBnoether,fdirac}. In
other words, the existence of the symmetry is connected to the existence of a vector field

\begin{eqnarray} 
X= \alpha^i (q)\frac{\partial}{\partial q^i}+\frac{d\alpha^i(q)}{dt}\frac{\partial}{\partial {\dot q}^i}\,,
\end{eqnarray}
where at least one of the components  $\alpha^i(q)$ have to be different from zero to generate a symmetry. In our case,
the generator of symmetry is 

\begin{eqnarray} 
X=\alpha \frac{\partial}{\partial a}+ \beta\frac{\partial}{\partial R}+\gamma  \frac{\partial}{\partial T}+{\dot \alpha} \frac{\partial}{\partial  \dot a}+  {\dot \beta}\frac{\partial}{\partial \dot R}+{\dot \gamma}  \frac{\partial}{\partial\dot T}\,.
\label{ourX}
\end{eqnarray}
The functions $\alpha, \beta, \gamma$ depend on the variables $a, R, T$ and then
\begin{eqnarray} 
{\dot \alpha}\,=\, \frac{\partial \alpha}{\partial a}{\dot a}+\frac{\partial \alpha}{\partial R}{\dot R}+\frac{\partial \alpha}{\partial T}{\dot T}\,,\quad
{\dot \beta}\,=\, \frac{\partial \beta}{\partial a}{\dot a}+\frac{\partial \beta}{\partial R}{\dot R}+\frac{\partial \beta}{\partial T}{\dot T}\,,\quad
{\dot \gamma}\,=\, \frac{\partial \gamma}{\partial a}{\dot a}+\frac{\partial \gamma}{\partial R}{\dot R}+\frac{\partial \gamma}{\partial T}{\dot T}\,.
\end{eqnarray}
As stated above, a  Noether symmetry exists if at
least one of them is different from zero.  Their analytic forms can be found by making explicit  Eq. (\ref{LX}), which corresponds to a set of partial differential equations given by
equating to zero the terms in ${\dot a}^2$,${\dot a}{\dot T}$, ${\dot a}{\dot R}$, ${\dot T}^2$, ${\dot R}^2$,${\dot R}{\dot T}$  and so on. In our specific case, we get a system of 7 partial differential
equations related to the fact that being the minisuperpace 3-dim, it is $1+n(n+1)/2$ as shown in \cite{cimento}. We have

\begin{eqnarray}\label{dota2}
&&6\alpha \left[\frac{\partial F(R,T)}{\partial R}-\frac{\partial F(R,T)}{\partial T}\right]+6\beta a  \left[\frac{\partial^2 F(R,T)}{\partial R^2}-\frac{\partial^2 F(R,T)}{\partial R\partial T}\right]+6 \gamma a\left[\frac{\partial^2 F(R,T)}{\partial R\partial T}-\frac{\partial^2 F(R,T)}{\partial T^2}\right]\nonumber\\&&+12 a \frac{\partial \alpha}{\partial a}\frac{\partial F(R,T)}{\partial R}-12 a\frac{\partial \alpha}{\partial a}\frac{\partial F(R,T)}{\partial T}+6a^2 \frac{\partial \beta}{\partial a}\frac{\partial^2 F(R,T)}{\partial R^2}+6a^2 \frac{\partial \gamma}{\partial a}\frac{\partial^2 F(R,T)}{\partial R\partial T}=0\,,
\end{eqnarray}

\begin{eqnarray}\label{dotadotT}
&&12\alpha a \frac{\partial^2 F(R,T)}{\partial R\partial T}+6\beta a^2 \frac{\partial^3 F(R,T)}{\partial R^2\partial T}+6\gamma a^2 \frac{\partial^3 F(R,T)}{\partial R\partial T^2}+6a^2  \frac{\partial \alpha}{\partial a} \frac{\partial^2 F(R,T)}{\partial R\partial T}\nonumber\\&&+12a\frac{\partial \alpha}{\partial T} \frac{\partial F(R,T)}{\partial R}-12a\frac{\partial \alpha}{\partial T} \frac{\partial F(R,T)}{\partial T}+6a^2\frac{\partial \beta}{\partial T} \frac{\partial^2 F(R,T)}{\partial R^2}  +6a^2 \frac{\partial \gamma}{\partial T}  \frac{\partial^2 F(R,T)}{\partial R\partial T}=0\,,
 \end{eqnarray}
 
  \begin{eqnarray}\label{dotadotR} 
  &&12 \alpha a \frac{\partial^2 F(R,T)}{\partial R^2} +6\beta a^2  \frac{\partial^3 F(R,T)}{\partial R^3}+6\gamma a^2 \frac{\partial^3 F(R,T)}{\partial R^2\partial T}+6a^2  \frac{\partial\alpha}{\partial a}\frac{\partial^2 F(R,T)}{\partial R^2} +12a  \frac{\partial\alpha}{\partial R}\frac{\partial F(R,T)}{\partial R}\nonumber\\&&-12 a  \frac{\partial\alpha}{\partial R} \frac{\partial F(R,T)}{\partial T}+6a^2   \frac{\partial\beta}{\partial R} \frac{\partial^2 F(R,T)}{\partial R^2}+6a^2 \frac{\partial\gamma}{\partial R}\frac{\partial^2 F(R,T)}{\partial R\partial T}=0\,,
   \end{eqnarray} 
  
   \begin{eqnarray}\label{dotT2}  
   6a    \frac{\partial\alpha}{\partial T} \frac{\partial^2F(R,T)}{\partial R\partial T}=0\,,
       \end{eqnarray} 
 
   \begin{eqnarray}\label{dotR2}  
  6a^2       \frac{\partial\alpha}{\partial R}\frac{\partial^2F(R,T)}{\partial R^2}=0
       \end{eqnarray}        
          \begin{eqnarray}\label{dotRdotT} 
          6a^2  \frac{\partial\alpha}{\partial R}  \frac{\partial^2F(R,T)}{\partial R\partial T}+6a^2  \frac{\partial\alpha}{\partial R} \frac{\partial^2F(R,T)}{\partial R^2}=0\,,
            \end{eqnarray}

  \begin{eqnarray}\label{senzadot}    
&&3\alpha a^2 \left[F(R,T)-T \frac{\partial F(R,T)}{\partial T}-R \frac{\partial F(R,T)}{\partial R}\right]-\beta a^3 \left[T \frac{\partial^2 F(R,T)}{\partial R\partial T}+ R \frac{\partial^2 F(R,T)}{\partial R^2}\right]\nonumber\\&&-\gamma a^3 \left[T \frac{\partial^2 F(R,T)}{\partial T^2}+R \frac{\partial^2 F(R,T)}{\partial R\partial T}\right]=0\,.
  \end{eqnarray}      
The above system is overdetermined and, if solvable, enables one to assign $\alpha,\beta,\gamma$ and $F(R,T)$.
The analytic form of $F(R,T)$ can be fixed by imposing, in the last equation of system (\ref{senzadot}), the conditions

\begin{equation}\label{sistesenza}
\left\{\begin{array}{ll}
F(R,T)-T \frac{\partial F(R,T)}{\partial T}-R \frac{\partial F(R,T)}{\partial R}=0\\
T \frac{\partial^2 F(R,T)}{\partial R\partial T}+ R \frac{\partial^2 F(R,T)}{\partial R^2}=0\\
T \frac{\partial^2 F(R,T)}{\partial T^2}+R \frac{\partial^2 F(R,T)}{\partial R\partial T}=0\\
\end{array}\right.\end{equation}
where the second and third equations  are  symmetric. However, it is clear that this is nothing else but an arbitrary choice since more general conditions are possible.  In particular, we can choose the functional forms:

\begin{eqnarray} 
F(R,T) = f(R)+ f(T)\,, \qquad F(R,T) = f(R) f(T)\,,
\end{eqnarray} 
from which it is easy to prove that the functional forms compatible with the system (\ref{sistesenza}) are:

\begin{eqnarray} 
F(R,T)= F_0 R+ F_1 T\,,\qquad 
F(R,T)= F_0 R^n T^{1-n}\,.
\end{eqnarray}
The first case is nothing else but the GR, the second gives interesting cases of possible extended theories as soon as $n\neq1$.

\subsection{The case $n=2$}
For  $n=2$,  the canonical  Lagrangian (\ref{PointLagra}) assumes the  form
\begin{eqnarray}\label{lagrasoln2}
{\cal L}=6 a^2 {\dot a} \left(\frac{2 {\dot R}}{T}-\frac{2 R {\dot T}}{T^2}\right)+6 a {\dot a}^2 \left(\frac{R^2}{T^2}+\frac{2
   R}{T}\right)
\end{eqnarray}
We can choose the variable ${\displaystyle \frac{R}{T}= \zeta}$ so reduce the system. The above  Lagrangian is transformed into
\begin{eqnarray}\label{lagraredun2}
{\cal L}= 2 a^2 {\dot a} {\dot \zeta}+ 2 a {\dot a}^2\zeta+ a{\dot a}^2 \zeta^2
\end{eqnarray}
Clearly we have reduced the dynamics  assuming that $\zeta$  depends on $R$ and $T$.
The Euler-Lagrange equations are 
\begin{eqnarray}
&& {\ddot \zeta}
+\left(\frac{\dot a}{a}\right)^2 \zeta
+2\left(\frac{\ddot a}{a} \right)\zeta+2\left( \frac{\dot a }{a}\right) {\dot \zeta}+\frac{1}{2} \left(\frac{\dot a}{a}\right)^2 \zeta^2+ \left(\frac{\ddot a }{a}\right)\zeta^2 + 2 \left( \frac{\dot a }{a}\right)\zeta {\dot \zeta} =0\,,
\label{motoz1a}\\
&& \left(\frac{\dot a}{a}\right)^2 + \frac{\ddot a}{a} - \left(\frac{\dot a}{a}\right)^2 \zeta =0\,,
\end{eqnarray} 
and the energy condition 
\begin{eqnarray}
 \left(\frac{\dot a}{a}\right)^2 \zeta^2+ 2  \left(\frac{\dot a}{a}\right)^2 \zeta+ 2\left( \frac{\dot a}{a}\right) {\dot \zeta}=0\label{energyz}\,.
\end{eqnarray} 
Clearly we lost an equation of motion because the relation between the two variables $R$ and $T$ is fixed by $\zeta$.  Immediately, an exact solution is 
\begin{equation}
a(t)=a_0 t^{1/2}\,,\qquad \zeta=0\,.
\end{equation}
which is a radiation solution. Another solution is achieved for $\zeta=1$ but it is a trivial one being $a(t)=a_0$. This means that these two solutions, in the case $n=2$, are quite natural due to the fact that the asymptotic behavior of $R$ is $1/t^2$ like that of $T$ that it is always $\sim 1/t^2$. Then $\zeta$ can be either equal to zero or equal to a constant. 

\section{Conclusions}
\label{cinque}  
  We have considered the Noether Symmetry Approach for  cosmology coming from a generalized gravitational theory $F(R,T)$ which is a
function of   the torsion scalar $T$ and of the Ricci curvature scalar $R$. The  existence of the Noether symmetry selects suitable $F(R,T)$ models and allows to reduce dynamics.  As a consequence, the reduction process  allows to achieve exact solutions. We have used   Lagrange multipliers to derive a point-like canonical Lagrangian. In this sense, the functions $a,T,R$ can be considered as independent fields \cite{makarenko}.  In a forthcoming paper, we will full develop the method addressing physically observable models.


\end{document}